\documentclass[conference]{IEEEtran}
\IEEEoverridecommandlockouts
\usepackage{cite}
\usepackage{amsmath,amssymb,amsfonts,bm}
\usepackage{graphicx}
\usepackage{textcomp}
\usepackage{xcolor}
\usepackage{comment}
\usepackage{amsmath,amssymb,amsfonts}
\usepackage{tikz}
\usetikzlibrary{external}
\usetikzlibrary{arrows,automata,positioning}
\usetikzlibrary{arrows.meta}
\usepackage{graphicx}
\usepackage{textcomp}
\usepackage{xcolor}
\usepackage{lipsum}
\usepackage{booktabs}
\usepackage{color, colortbl}
\usepackage{algorithm}
\usepackage{algpseudocode}
\usepackage{dsfont}
\usepackage[acronyms]{glossaries}
\usepackage{tcolorbox}
\usetikzlibrary{calc, positioning, arrows.meta, fit}

\def\BibTeX{{\rm B\kern-.05em{\sc i\kern-.025em b}\kern-.08em
    T\kern-.1667em\lower.7ex\hbox{E}\kern-.125emX}}

\pgfdeclarelayer{lay00}
\pgfdeclarelayer{back00}
\pgfdeclarelayer{lay01}
\pgfdeclarelayer{back01}
\pgfdeclarelayer{lay02}
\pgfdeclarelayer{back02}
\pgfdeclarelayer{lay03}
\pgfdeclarelayer{back03}
\pgfdeclarelayer{back04}
\pgfdeclarelayer{lay10}
\pgfdeclarelayer{back10}
\pgfdeclarelayer{lay11}
\pgfdeclarelayer{back11}
\pgfdeclarelayer{lay12}
\pgfdeclarelayer{back12}
\pgfdeclarelayer{lay13}
\pgfdeclarelayer{back13}
\pgfdeclarelayer{back14}
\pgfdeclarelayer{lay30}
\pgfdeclarelayer{back30}
\pgfdeclarelayer{lay31}
\pgfdeclarelayer{back31}
\pgfdeclarelayer{lay32}
\pgfdeclarelayer{back32}
\pgfdeclarelayer{lay33}
\pgfdeclarelayer{back33}
\pgfdeclarelayer{back34}

\pgfsetlayers{back34,back33,lay33,back32,lay32,back31,lay31,back30,lay30,back14,back13,lay13,back12,lay12,back11,lay11,back10,lay10,back04,back03,lay03,back02,lay02,back01,lay01,back00,lay00,main}

\newlength{\figSpaceX}
\newlength{\figSpaceY}
\newlength{\nodeSize}
\colorlet{group color}{white}
\colorlet{components color}{black!15!white}

\newacronym{rcs}{RCS}{radar cross section}
\newacronym{dtt}{DTT}{detect-then-track}
\newacronym{mot}{MOT}{multiobject tracking}
\newacronym{tbd}{TBD}{track-before-detect}
\newacronym{rfs}{RFS}{random finite set}
\newacronym{bp}{BP}{belief propagation}
\newacronym{pdf}{PDF}{probability density function}
\newacronym{snr}{SNR}{signal-to-noise ratio}
\newacronym{sbl}{SBL}{sparse Bayesian learning}
\newacronym{ospa}{OSPA}{optimal sub-pattern assignment}
\newacronym{vmp}{VMP}{variational message passing}
\newacronym{slam}{SLAM}{simultaneous localization and mapping}
\newacronym{po}{PO}{potential objects}

\newcommand{\trans}{\text{T}}

\newcommand{\ist}{\hspace*{.3mm}}
\newcommand{\rmv}{\hspace*{-.3mm}}
\newcommand{\iist}{\hspace*{1mm}}

\newcommand{\nn}{\nonumber}

\begin{document}

\title{\huge Variational Message Passing-based Multiobject Tracking for MIMO-Radars using Raw Sensor Signals\\

\thanks{
This work is partly funded by the Thomas B. Thriges Foundation grant 7538-1806.}
}

\author{
    \IEEEauthorblockN{Anders Malthe Westerkam$^\dagger$, Jakob M\"oderl$^*$, Erik Leitinger$^*$, and Troels Pedersen$^\dagger$
    }
    \IEEEauthorblockA{$\dagger$Aalborg University, Aalborg Denmark. Email: \{amw troels\}@es.aau.dk}
    \IEEEauthorblockA{$*$Graz University of Technology, Graz, Austria. Email: \{jakob.moderl, erik.leitinger\}@tugraz.at}
}

\maketitle

\begin{abstract}
In this paper, we propose a direct \gls{mot} approach for MIMO-radar signals that operates on raw sensor data via \gls{vmp}. Unlike classical \gls{tbd} methods, which often rely on simplified likelihood models and exclude nuisance parameters (e.g., object amplitudes, noise variance), our method adopts a superimposed signal model and employs a mean-field approximation to jointly estimate both object existence and object states. By considering correlations within in the radar signal due to closely spaced objects and jointly estimating nuisance parameters, the proposed method achieves robust performance for 
close-by objects and in low-\gls{snr} regimes. Our numerical evaluation based on MIMO-radar signals demonstrate that our \gls{vmp}-based direct-\gls{mot} method outperforms a \gls{dtt} pipeline comprising a super-resolution 
\gls{sbl}-based estimation stage followed by classical \gls{mot} using global nearest neighbor data association and a Kalman filter.
\end{abstract}
\begin{IEEEkeywords}
Multiobject tracking, Track-Before-Detect, Direct Tracking, Variational Message Passing
\end{IEEEkeywords}
\glsresetall
\section{Introduction}
In recent years unmanned aerial vehicles or drones have become a bigger part of both private, commercial, and military applications. This means that it is easier than ever to gain access to the airspace with the cost of drones continuing to go down furthermore modern drone systems even allows multiple drones to be operated by a single user. All this puts restricted airspaces, such as around airports, under more risk of getting penetrated either accidentally or by adversarial operators. To detect such penetrations radars have been used for many years as they are reliable in any light and weather condition \cite{Richards2014}. However the detection of drones using radar is complicated by their make and small size which results in poor reflective properties normally expressed through a small \gls{rcs}. in addition drones are normally slow moving with respect to the surrounding clutter and hence normal high pass filtering in Doppler may not isolate the drone signature \cite{Poitevin2017,Gong2023,Quevedo2019}. This necessitates the development of algorithms to detect, localize and track multiple slow moving weak objects.

In radar signal processing, \gls{mot} (i.e. the tasks of detecting, localizing, and tracking multiple objects) has traditionally been carried out sequentially via \gls{dtt} algorithms, which rely on pre-processed object estimates--i.e., measurements--rather than raw radar signals \cite{Wei2017}. The \gls{dtt} appraoch can infer the states of objects from measurements provided by one or more sensors, even when the number of objects is unknown \cite{BarWilTia:B11,Mahler2007, MeyerProc2018, LiLeiVenTuf:TWC2022,VenLeiTerMeyWit:TWC2024, Zhang2024}. However, \gls{dtt}-\gls{mot} methods frequently exhibit suboptimal performance when tracking weak objects, since such objects may be masked by clutter or noise \cite{Ristic2020}, leading to missed detections and, consequently, no information being passed to the tracker.

To address this shortcoming, \gls{tbd} methods have been developed to operate directly on received radar signals, rather than on intermediate measurements, thereby improving tracking performance in scenarios involving weak objects \cite{TonSha:TAES1998, Rutten2005, Davey2007, MoySpaLam:TAES2011, PapVoVoFanBea:TAES2015, LepRabLeG:TAES2016, Ristic2020 ,Lehmann2012, Zhichao2020}.  
Many different \gls{tbd} approaches have emerged, including batch-processing techniques based on maximum likelihood estimation \cite{TonSha:TAES1998},  the Hough transform \cite{MoySpaLam:TAES2011} and dynamic
programming \cite{Barniv:TAES1985}; however, their computational complexity is typically too high for real-time operation. In contrast, well-established real-time Bayesian \gls{tbd} \gls{mot} methods for an unknown number of objects can be broadly divided into two categories: those using \gls{rfs} filters \cite{Kropfreiter2024,Ristic2020,KimRisGuaRos:TAES2021}, and those based on message passing on factor graphs via \gls{bp} \cite{Liang2023}, which exploit the natural factorization of the posterior \gls{pdf} to achieve high scalability. In \cite{Liang2023}, a \gls{bp}-based approach is proposed that uses a cell grid potentially containing objects or random noise. This method accounts for objects contributing to multiple cells and employs a birth-death model to facilitate the initialization and termination of tracks, demonstrating robustness in weak-object scenarios. Most commonly used \gls{tbd} methods are based on simplified likelihood models that do not fully capture the true radar signal. In particular, they typically use a point-spread function, assume that nuisance parameters (e.g. amplitudes and noise variance) are known \cite{Ristic2020,PapVoVoFanBea:TAES2015}, use matched-filtered radar signals \cite{Lehmann2012}, and impose a separability condition on the likelihood function (i.e, samples are treated as independent and only one object is allowed per sample) \cite{Ristic2020,KroWilMey:FUSION2021,Kropfreiter2024}. In \cite{LepRabLeG:TAES2016}, several more general likelihood models that directly include the radar signal have been proposed to relax these simplifying assumptions. In \cite{LiaLeiMey:Asilomar2023,LiaLeiMey:TSP2024}, a direct-multipath-based \gls{slam} method--where multipath-based \gls{slam} is strongly related to \gls{mot}--based on \gls{bp} message passing for superposition measurement models was introduced, operating directly on the full radar signal model. It is a \gls{tbd}-related approach, but unlike traditional \gls{tbd} methods, as a direct approach \cite{BiaRapWei:TVT2013}, it incorporates unknown nuisance parameters (e.g., amplitudes, object \gls{snr}, noise variance) and captures correlations in the raw sensor signal caused by closely spaced objects.

Other \gls{mot} methods employing message passing rely on a variational Bayesian framework, i.e., \gls{vmp}, but still operate on pre-processed measurements (i.e., apply a \gls{dtt} approach) \cite{LundgrenTSP2016, GanLiGod:TAES2024, BaiLanWanPanHaoLi:TAES2024}. A well-established real-time variational \gls{tbd} \gls{mot} method is the Histogram Probabilistic Multi-Hypothesis Tracker (HPMHT) \cite{Davey2007,DavWieVu:JSTSP2013}, which implements the expectation-maximization algorithm. However, tuning HPMHT parameters is reported to be challenging \cite{KimRisGuaRos:TAES2021}. In our previous work \cite{Westerkam2023}, involving some of the same authors, we proposed a direct-\gls{mot} method based on \gls{vmp} that also operates directly on the radar signal for joint localization and tracking of low-\gls{snr} objects. However, that work assumed a known number of objects and did not consider object detection or track formation.

In this paper, we present a direct-\gls{mot} method based on \gls{vmp} that accounts for an unknown number of objects and integrates both object detection and track formation. Specifically, our method employs a mean-field approximation and, in accordance with \cite{LiaLeiMey:Asilomar2023, LiaLeiMey:TSP2024}, is built on a superimposed signal model that captures correlations in raw sensor signals caused by closely spaced objects. Inspired by \cite{Badiu2017,Moederl2024}, this formulation enables the simultaneous estimation of object existence--modeled by a binary random variable--and individual object states (position, velocity, and potentially other kinematic parameters). The main contributions of this paper are as follows.
\begin{itemize}
    \item  We introduce a novel direct-\gls{mot} method based on \gls{vmp} to jointly estimate the number of objects and their individual states. 
    \item We derive closed-form message updates that effectively consider correlations within the raw data and remain computationally efficient by exploiting a mean-field approximation.
    \item We demonstrate using  MIMO radar signals that our method outperforms a \gls{dtt} approach consisting of a super-resolution \gls{sbl}-based estimation stage \cite{HansenSAM2014,GreLeiWitFle:TWC2024,Moederl2024} followed by classical \gls{mot} using global nearest neighbor data association and a Kalman filter for tracking.
\end{itemize}

\section{MIMO Radar Signal Model}

\begin{figure}[htbp]
\centerline{\includegraphics[width = 1 \linewidth]{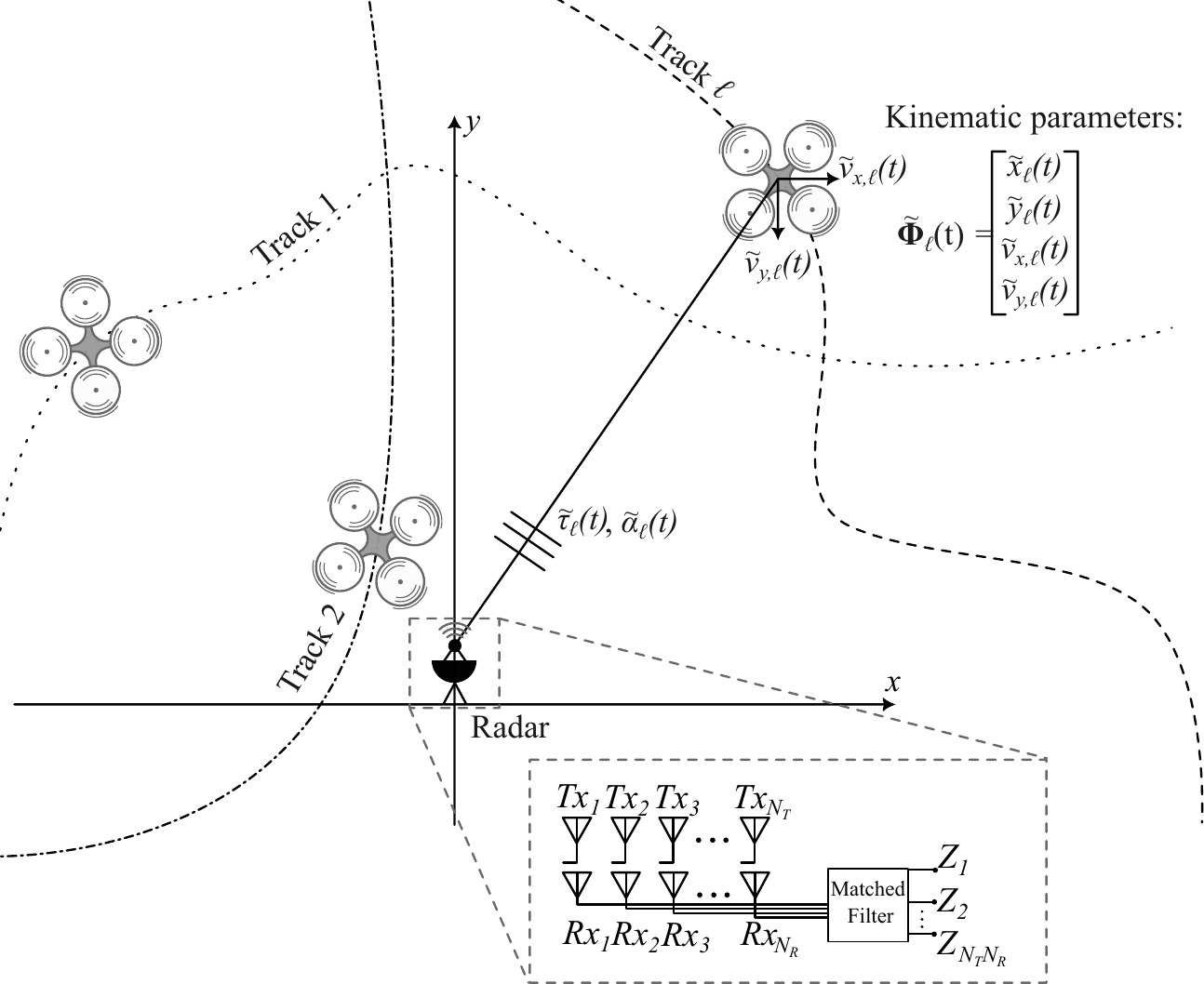}}
\caption{Scenario with an unknown number of objects $L_n$, each with their own kinematic parameters $\tilde{\bm{\Phi}}_{\ell,n}$ and reflectivity $\tilde{\alpha}_{\ell,n}$, being observed by a single $N_\text{T}\times N_\text{R}$ MIMO Radar.}
\label{fig:Overveiw}
\end{figure}

We consider a scenario with $L(t)$ objects in a clutter-free environment as exemplified in Fig.~\ref{fig:Overveiw}. Referring to a coordinate system with origin at the radar, each object $\ell$ is characterize by its state-space parameters $\tilde{\bm{\Phi}}_{\ell}(t) = [\bm{p}_{\ell}(t)^\trans \iist \bm{v}_{\ell}(t)^\trans]^\trans$, describing its position $\bm{p}_{\ell}(t) = [\tilde{x}_{\ell}(t)\iist \tilde{y}_{\ell}(t)]^\trans$ and velocity $\bm{v}_{\ell}(t) = [\tilde{v}_{\text{x},\ell}(t)\iist \tilde{v}_{\text{y},\ell}(t)]^\trans$, and by its radar cross section (RCS) $\sigma_{\text{R},\ell}$. The problem considered is to estimate both $L(t)$, the number of objects, and their respective states $\tilde{\bm{\Phi}}_{\ell}(t)$.

The MIMO radar is monostatic  with $N_\text{T}$ transmitters and $N_\text{R}$ receivers, where each antenna is assumed isotropic. Each transmitter $m$ emits a baseband signal $u_m(t)$ at carrier frequency $f_{c}$, and the signals from different transmitters are assumed mutually orthogonal. One MIMO pulse includes transmissions from all $N_\text{T}$ transmitters, requiring a total time of $T_{\mathrm{Tx}}\,N_\text{T}$. The time between consecutive MIMO pulses is $\Delta t$, giving a pulse repetition frequency (PRF) $\mathrm{PRF} = 1/\Delta t$. All objects in the radar’s field of view (FOV) reflect each pulse. Assuming slow object motion relative to the PRF, a “stop-and-hop” model is used:  the number of objects and the kinematic parameters of the $\ell$-th object remain constant during the $n$-th interval, i.e., $\tilde{\bm{\Phi}}_{\ell}(t) = \tilde{\bm{\Phi}}_{\ell,n}$ and $L(t)=L_n$ for $n \Delta t \leq t \leq (n+1)\Delta t$. Since velocities are low, Doppler effects are neglected. Only the direct path is considered, and all objects lie in the far field. After down-conversion, the baseband signal at the $j$-th receiver is
\vspace{-2mm}
\begin{align}\label{eq:signal_model_time}
    y_{j,n}(t) = \sum_{\ell=1}^{L_n} \sum_{m=1}^{N_\text{T}} \tilde{\alpha}_{\ell,n} \ist a_{j,m}(\tilde{\theta}_{\ell,n}) \ist u_m\bigl(t - \tilde{\tau}_{\ell,n}\bigr) + w_j(t)  \\[-6mm]\nn 
\end{align}
where $\tilde{\alpha}_{\ell,n} = \sqrt{\sigma_{\mathrm{R},\ell}} \, e^{i\,2\pi\,f_{c}\,\tau_{\ell,n}}$ is the complex weight of the $\ell$-th object. The array steering function $a_{j,m}(\tilde{\theta}_{\ell,n})$ at receiver $j$ for transmitter $m$ is evaluated at the bearing $\tilde{\theta}_{\ell,n} = \arctan(x_{\ell,n} / y_{\ell,n})$. The term $u_m\bigl(t - \tilde{\tau}_{\ell,n}\bigr)$ is the transmitted baseband signal from transmitter $m$ delayed by $\tilde{\tau}_{\ell,n}$, where $\tilde{\tau}_{\ell,n} = 2\,\sqrt{(x_{\ell,n})^2 + (y_{\ell,n})^2} / c$ represents the two-way propagation time for the $\ell$-th object, and $c$ is the speed of light. The last term $w_j(t)$ denotes white, complex, circularly-symmetric Gaussian noise with variance $\sigma_w^2$.

\section{System Model}

\subsection{State Vectors and Inference Signal Model}

We consider $K$ \gls{po}  with $K > L_n$ and time-varying states.
The $k$-th PO at time $n$ has state $\bm{\Phi}_{n,k} = [\bm{p}^\trans_{k,n} \iist \bm{v}^\trans_{k,n}]^\trans$ and complex-valued reflectivity $\alpha_{k,n}$. Each PO is also associated with a binary existence indicator $\xi_{n,k}\in \{0,1\}$, meaning that the PO is present if and only if $\xi_{n,k} = 1$ {such that $L_n=\sum_{k=1}^{K}\xi_{n,k}$}. After sampling and matched filtering in the frequency domain, the received signal $\bm{Z}_n \in \mathbb{C}^{N \times 1}$ with $N_Z = N_s N_\text{T} N_\text{R}$ is\footnote{Note that the maximum number of \gls{po} is given by the signal samples, i.e., $K_\text{max}=N_Z$.}
\vspace{-2mm}
\begin{align}\label{eq:Z_constructio}
    \bm{Z}_n 
    = \sum_{k=1}^{K} \alpha_{k,n}\, \xi_{n,k} \, \bm{S}\bigl(\bm{\Phi}_{n,k}\bigr) + \tilde{\bm{W}}
    \\[-6mm]\nonumber
\end{align}
is the matched-filter output, and $\tilde{\bm{W}} \in \mathbb{C}^{N \times 1}$ is colored noise with zero mean and covariance $\bm{\Lambda}_Z$. The term 
\vspace{-2mm}
\begin{align}
   \bm{S}\bigl(\bm{\Phi}_{n,k}\bigr) = \sum_{m=1}^{N_\text{T}} \bm{A}_m\bigl(\theta_{n,k}\bigr)\otimes \bm{h}_m(\tau_{n,k})\\[-6mm] \nonumber
\end{align}
represents the spatio-temporal steering vector from the $k$-th PO at time $n$. Here, $\bm{a}_m(\theta_{n,k}) \in \mathbb{C}^{N_\text{R} \times 1}$ is the receive steering vector for bearing $\theta_{n,k}$, $\bm{h}_m(\tau_{n,k})\in\mathbb{C}^{N_\text{T} N_s \times 1}$ is the matched-filtered transmit-signal spectrum for delay $\tau_{n,k}$, and $\otimes$ denotes the Kronecker product.

\subsection{Probabilistic Model}

The existence indicator $\xi_{k,n}$, the object state $\bm{\Phi}_{k,n}$, and complex-valued reflectivities $\alpha_{k,n}$, process-noise covariances matrices $\bm{\Lambda}_{k,\text{a}}$ are considered unknown and time-varying. We define the stacked vectors 
$\bm{\Phi}_n \triangleq [\bm{\Phi}_{1,n}^{\trans}, \dots, \bm{\Phi}_{K,n}^{\trans}]^{\trans}$,
$\bm{\xi}_n \triangleq [\xi_{1,n}, \dots, \xi_{K,n}]^{\trans}$,
$\bm{\alpha}_n \triangleq [\,\alpha_{1,n},\ist \cdots, \ist\,\alpha_{K,n}\,]^{\trans}$,
$\bm{\Lambda}_a \triangleq [\bm{\Lambda}_{\text{a},1}, \dots, \bm{\Lambda}_{\text{a},K}]$,
and similarly collect measurements $\bm{Z}_n$. Let 
$\bm{\Phi}_{0:N} \triangleq [\bm{\Phi}_0,\ist \cdots, \ist\bm{\Phi}_N]$,
$\bm{\xi}_{0:N} \triangleq [\bm{\xi}_0,\ist \cdots, \ist\bm{\xi}_N]$,
$\bm{\alpha}_{0:N} \triangleq [\bm{\alpha}_0,\ist \cdots, \ist\bm{\alpha}_N]$, and
$\bm{Z}_{0:N} \triangleq [\bm{Z}_0,\ist \cdots, \ist\bm{Z}_N]$. {Where $N$ refers  to the last recorded time instance, and will hence grow as more data is collected}

\subsubsection{State Transition Model}

We assume that the \gls{po} state evolve independently across $k$ and $n$ and the joint state-transition \gls{pdf} factorizes as 
\begin{equation}\label{eq:SST}
    p(\bm{\Phi}_{n} \mid \bm{\Phi}_{0:n-1},\,\bm{\Lambda}_{\text{a}}) = \prod_{n=1}^N \ist \prod_{k=1}^K p(\bm{\Phi}_{k,n} \mid \bm{\Phi}_{k,n-1},\,\bm{\Lambda}_{\text{a},k})
\end{equation}
where $p(\bm{\Phi}_{k,n} \mid \bm{\Phi}_{k,n-1},\,\bm{\Lambda}_{\text{a},k})$ follows a first-order Markov process with linear dynamics. In particular,
\begin{equation}\label{eq:phi-markov}
  \bm{\Phi}_{k,n} 
  \;=\; 
  \bm{T}\,\bm{\Phi}_{k,n-1} + \bm{G}\,\bm{a}_k
\end{equation}
where $\bm{a}_k$ is a zero-mean Gaussian random vector with covariance $\bm{\Lambda}_{k,\text{a}}$, i.e., 
$\bm{a}\sim \mathcal{N}(\bm{0},\,\bm{\Lambda}_{k,\text{a}})$. The matrices $\bm{T}$ and $\bm{G}$ are known, for instance with a constant-velocity motion model one might have
\[
  \bm{T} 
  = 
  \begin{bmatrix}
    1 & 0 & \Delta t & 0 \\
    0 & 1 & 0       & \Delta t \\
    0 & 0 & 1       & 0 \\
    0 & 0 & 0       & 1
  \end{bmatrix},
  \quad
  \bm{G} 
  = 
  \begin{bmatrix}
    \tfrac{\Delta t^2}{2} & 0 & 0 & 0 \\
    0 & \tfrac{\Delta t^2}{2} & 0 & 0 \\
    0 & 0 & \Delta t & 0 \\
    0 & 0 & 0 & \Delta t
  \end{bmatrix}.
\]

\subsubsection{Evolution of Existence Indicator}

We model $\xi_{k,n}$ as a birth-death process independent across $k$, i.e.,
\begin{align}
    p\bigl(\xi_{k,n} \mid \xi_{k,n-1}\bigr) =
	\begin{cases}
		p_\text{s}, 	&\xi_{k,n}= 1,\ \xi_{k,n-1}=1\\
		1 - p_\text{b}, &\xi_{k,n} = 0,\ \xi_{k,n-1}=1 \\        
		p_\text{b}, 	&\xi_{k,n} = 1,\ \xi_{k,n-1}=0 \\
		1 - p_\text{s}, &\xi_{k,n}= 0,\ \xi_{k,n-1}=0  
	\end{cases}
	\label{eq:xi-markov1} \\[-6mm]\nn
\end{align}

\subsubsection{Reflectivity Model}

In modeling the distribution of \(\alpha_{k,n}\), it is well known that an object's return strength can vary significantly from one time step to the next \cite{LepRabLeG:TAES2016}. Consequently, we assume \(\alpha_{k,n}\) to be a priori independent across both time \(n\) and PO index \(k\). They are treated as nuisance parameters, with the prior PDF
\vspace{-2mm}
\begin{align}\label{eq:priorweigths}
  p(\boldsymbol{\alpha}_{n}) = \prod_{k=1}^K\mathrm{CN}\ist\bigl(\alpha_{k,n} \,;\, 0,\,\lambda_{\alpha,\text{p}}^{-1}\bigr)\\[-5mm]\nn
\end{align}
with precision $\lambda_{\alpha,\text{p}}$, where $\mathrm{CN}(\bm{x};\bm{\mu},\bm{\Sigma})$ denotes the \gls{pdf} of a multi-variate complex-Gaussian distribution of the variable $\bm{x}$ with mean $\bm{\mu}$ and covariance $\bm{\Sigma}$
\subsubsection{Observation Model}

At each time $n$, a measurement vector $\bm{Z}_n$ is observed. The likelihood function is given by
\vspace{-2mm}
\begin{align}
  &p\bigl(\bm{Z}_n \mid \bm{\Phi}_n,\,\bm{\xi}_n,\,\bm{\alpha}_n\bigr)\nn\\
  &\hspace{8mm} = \mathrm{CN}\ist\Bigl(
    \bm{Z}_n;
    \sum_{k=1}^{K} \xi_{k,n}\,\alpha_{k,n}\ist\bm{S}(\bm{\Phi}_{k,n}),
    \bm{\Lambda}_Z
  \Bigr)\ist. \label{eq:obs-likelihood} \\[-7mm] \nn
\end{align}

\subsubsection{Joint Posterior PDF}

\tikzsetnextfilename{Baysian_Network}
\begin{figure}[t]
    \centering
    \includegraphics[width = 1\linewidth]{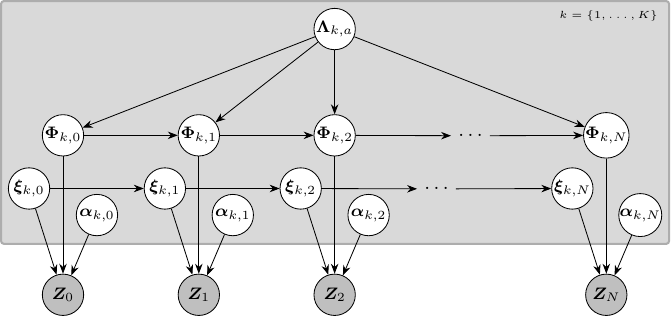}    
    \caption{%
      Bayesian network representation of the multi-object detection and tracking problem.
      Each object $k$ has state $\bm{\Phi}_{k,n}$, existence $\xi_{k,n}$, and reflectivity $\alpha_{k,n}$. 
      Shaded nodes denote the measured data $\bm{Z}_n$.
      }
    \label{fig:Baysian_graph}
\end{figure}

Using \eqref{eq:SST}, \eqref{eq:xi-markov1}, \eqref{eq:priorweigths}, and \eqref{eq:obs-likelihood}, the joint posterior PDF can be written as
\begin{align}
&p\bigl(\bm{\Phi}_{0:N},\,\bm{\xi}_{0:N},\,\bm{\alpha}_{0:N},\,\bm{\Lambda}_a,\,|\bm{Z}_{0:N}\bigr)\nonumber\\
  &\quad \propto p(\bm{\alpha}_0)\ist p(\bm{\xi}_0)\ist\biggl(\prod_{k'=1}^{K} p(\bm{\Phi}_{k',0})\biggr)
  \nonumber\\
  &\qquad \times
  \prod_{n=1}^{N}\biggl(
    p(\bm{Z}_n \mid \bm{\Phi}_n,\,\bm{\xi}_n,\,\bm{\alpha}_n)\,
    p(\bm{\alpha}_n)\,
    p(\bm{\xi}_n \mid \bm{\xi}_{n-1})\nonumber\\
  &\qquad \times
    \prod_{k=1}^{K}
      p(\bm{\Phi}_{k,n} \rmv\mid\rmv \bm{\Phi}_{k,n-1},\,\bm{\Lambda}_{\text{a},k}) \ist p(\bm{\Lambda}_{\text{a},k})\biggr)\label{eq:joint-factorization}
\end{align}
where $p(\bm{\alpha}_0)$, $p(\bm{\xi}_0)$, and $p(\bm{\Phi}_{k,0})$ represent the prior PDFs at time $n=0$ and $ p(\bm{\Lambda}_{\text{a},k})$ represents the prior \gls{pdf} on the process-noise covariance. 
Figure~\ref{fig:Baysian_graph} illustrates the corresponding Bayesian network, with shaded nodes representing the measured data $\bm{Z}_n$ and clear nodes representing the unknown random variables.

\section{Proposed Algorithm}

Based on the joint posterior PDF in \eqref{eq:joint-factorization}, our goal is to estimate the number of objects at each time step, ${L}_n$, by checking the number of existence indicators, $\xi_{k,n}$, that has a probability of being one above a certain threshold $\delta$, and then to estimate their associated, states \(\bm{\Phi}_{k,n}\) together with \(\alpha_{k,n}\) and \(\bm{\Lambda}_{\text{a},k}\).

\subsection{Mean-field VMP Approach}

Since the posterior \gls{pdf} \eqref{eq:joint-factorization} is intractable, we employ a structured mean-field approach. Specifically, we approximate \eqref{eq:joint-factorization} by a factorized proxy \gls{pdf} given by
\begin{equation}\label{eq:surogate_expression}
    q(\bm{\Phi}_n,\bm{\xi}_n,\bm{\alpha}_n) = \prod_{n=0}^{N}q(\bm{\alpha}_n)\prod_{k=1}^{K}q(\bm{\Phi}_{k,n})q(\xi_{k,n})q(\bm{\Lambda}_{k,a}).
\end{equation}
The optimal proxy \glspl{pdf} are determined by minimizing the Kullback-Leibler (KL) divergence between the posterior \gls{pdf} {\eqref{eq:joint-factorization}} and the proxy \gls{pdf} {\eqref{eq:surogate_expression}} equivalently to maximizing the evidence lower bound (ELBO)
\begin{equation}\label{eq:ELBO_main_article}
    \mathcal{L}(q) = \mathbb{E} \big[\ln{p\bigl(\bm{\Phi}_{0:N},\,\bm{\xi}_{0:N},\,\bm{\alpha}_{0:N},\,\bm{\Lambda}_a,\,|\bm{Z}_{0:N}\bigr)}\big] + H(q)
\end{equation}
where $H(\cdot)$ is the entropy, and the expectation is with respect to the proxy \gls{pdf} in \eqref{eq:surogate_expression}.

\subsubsection{Complex Reflectivity and Existence Indicator Updates} Starting by finding $q(\boldsymbol{\alpha}_n)$, and $q(\boldsymbol{\xi}_n)$, these marginals have a high interdependence and it was found that doing a joint optimization of the ELBO as in \cite{Moederl2024,Badiu2017}, yields the best result.
The derivation is found in App.~\ref{app:alpha_xi_optimization}, and results in,
\vspace{-2mm}
\begin{align}
        q(\bm{\alpha}_n) = \mathrm{CN}(\bm{\alpha}_n; \bm{\mu}_{\bm{\alpha},n},\bm{\Lambda}_{\bm{\alpha},n})\label{eq:update_alpha}\\[-6mm]\nn
\end{align}
with
\vspace{-2mm}
\begin{align}
    \bm{\Lambda}_{\bm{\alpha},n} &= \bm{M}_n\odot\mathbb{E}_{\bm{\Phi}}[\langle\bm{S}^T_n|\bm{\Lambda}_z|\bm{S}^T_n\rangle]+ \lambda_{a,p}\bm{I}\label{eq:Lambda_alpha}\\
    \bm{\mu}_{\alpha}^{(n)} &= \bm{\Lambda}_\alpha^{-1}\langle\bm{S}^T_n\bar{\bm{\xi}}|\Lambda_z|\bm{Z}_n\rangle\label{eq:mu_alpha}\\[-6mm]\nn
\end{align}
where $\langle\cdot|\cdot\rangle$ is the bra-ket notation for inner products, the expectation is taken using the delta method as shown in App.~\ref{app:div_of_KL_divergence}, and $\odot$ is the element wise multiplication. The matrices $\bm{M}_n$, $\bm{S_n}^T$, and $\bar{\bm{\xi}}$ are given as 
$\bm{S}_n^T = [\bm{S}(\bm{\Phi}_{1,n}) \ist\cdots\ist \bm{S}(\bm{\Phi}_{k,n})]$, $\bar{\bm{\xi}} = \text{diag}([\bar{\xi}_{1,n}\ist\cdots\ist \bar{\xi}_{k,n}]^T)$, and
\begin{align}
    \bm{M}_n &= \begin{bmatrix}
        \bar{\xi}_{1,n} & \bar{\xi}_{2,n}\bar{\xi}_{1,n} & \hdots & \bar{\xi}_{k,n}\bar{\xi}_{1,n} \\
        \bar{\xi}_{1,n}\bar{\xi}_{2,n} & \bar{\xi}_{2,n} & \hdots & \bar{\xi}_{k,n}\bar{\xi}_{2,n} \\
        \vdots & \vdots & \ddots & \vdots \\
        \bar{\xi}_{1,n}\bar{\xi}_{k,n} & \bar{\xi}_{2,n}\bar{\xi}_{k,n} & \hdots &\bar{\xi}_{k,n}
    \end{bmatrix}
\end{align}
where $\bar{\cdot}$ designates mean. The Bernoulli distribution, $q(\xi_{k,n})$, is fully defined by its mean which may be expressed as,
\begin{multline}\label{eq:update_xi}
     \hat{\bar{\xi}}_{k,n} = \underset{\bar{\xi}_{k,n}}{\text{argmax}}\phantom{m} \langle \bm{\mu}_{\alpha,n}|\bm{\Lambda}_{\alpha,n}|\bm{\mu}_{\alpha,n}\rangle - \ln{|\bm{\Lambda}_{\alpha,n}|} \\+ H(q(\xi_{k,n})) + \bar{\xi}_{k,n}g(\bar{\xi}_{k,n-1}).
\end{multline}
where $\text{logit}(x) = \ln{\frac{x}{1-x}}$, and
\begin{equation}\label{eq:g_def}
    g(\bar{\xi}_{k,n-1}) = \bar{\xi}_{k,n-1}(\text{logit}(p_s)-\text{logit}(p_b)) + \text{logit}(p_b).
\end{equation}

\subsubsection{\gls{po} State Update} For $q(\bm{\Phi}_{k,n})$, we follow the method outlined in \cite{bishop2007} (Chapter 10) which leads to a local optimum in the KL divergence and the surogate is expressed as,
\begin{multline}\label{eq:sorrogate_of_Phi_Bishop}
    \ln{q(\bm{\Phi}_{k,n})} = \mathbb{E}_{\backslash \bm{\Phi}_{k,n}}[p(\boldsymbol{Z}_n|\boldsymbol{\Phi}_n,\boldsymbol{\xi}_n,\boldsymbol{\alpha}_n)]\\+ \mathbb{E}_{\backslash \bm{\Phi}_{k,n}}[p(\bm{\Phi}_{k,n}|\bm{\Phi}_{k,n-1},\bm{\Lambda}_{k,a})] \\+ \mathbb{E}_{\backslash \bm{\Phi}_{k,n}}[p(\bm{\Phi}_{k,n+1}|\bm{\Phi}_{k,n},\bm{\Lambda}_{k,a})] + \text{C}
\end{multline} 
where $\mathbb{E}_{\backslash\bm{\Phi}_{k,n}}[\cdot]$ is the expectation with respect to the surogate in \eqref{eq:surogate_expression} excluding the variable $\bm{\Phi}_{k,n}$. Each term in \eqref{eq:sorrogate_of_Phi_Bishop} can be viewed as a message going to $\bm{\Phi}_{k,n}$, we denote these messages $\epsilon$. Starting with the first term $\epsilon^{(\bm{Z}_n\rightarrow\bm{\Phi}_{k,n})}$ we note, and as has been shown in previous work \cite{Kitchen2025}, that this message does not have a closed form expression. 
To obtain closed form solutions for all marginals, we restrict $\epsilon^{(\bm{Z}_n\rightarrow\bm{\Phi}_{k,n})}$ to be a Gaussian which minimizes the KL divergence w.r.t. the true message as the Gaussian is fully defined by its mean and covariance this is written as,
\begin{equation}\label{eq:KL_min_message}
    \{\bar{\bm{\epsilon}}_{k,g,n},\bar{\bar{\bm{\epsilon}}}_{k,g,n}\} = \underset{\bar{\bm{\epsilon}}_g,\bar{\bar{\bm{\epsilon}}}_g}{\text{argmin}} \, \mathcal{D}_{KL}({\epsilon}_g||{\epsilon}^{(\bm{Z}_n \rightarrow \bm{\Phi}_{k,n})}).
\end{equation}
Here ${\epsilon}_g$ is the Gaussian message, and $\bar{\bar{\cdot}}$ denotes the covariance matrix, the expression for the KL divergence is derived as \eqref{eq:KL_divergence} in App.~\ref{app:div_of_KL_divergence}.
The messages passed along the kinematic chain is,
\vspace{-2mm}
\begin{equation} 
    \epsilon^{(\bm{\Phi}_{k,n-1}\rightarrow\bm{\Phi}_{k,n})} = \mathrm{N}\left(\bm{\Phi}_{k,n};\bm{T}\bm{\Phi}_{k,n-1},\Check{\bar{\bm{\Lambda}}}_{k,a}\right)
\end{equation}
\begin{multline}
    \epsilon^{(\bm{\Phi}_{k,n+1}\rightarrow\bm{\Phi}_{k,n})} = \mathrm{N}\left(\bm{\Phi}_{k,n};\bm{T}^{-1}\bm{\Phi}_{n+1}^{(k)},\bm{T}^{T}\Check{\bar{\bm{\Lambda}}}_{k,a}\bm{T}\right)
\end{multline}
where $\Check{\bar{\bm{\Lambda}}}_{k,a} = \bm{G}^{-T}\bar{\bm{\Lambda}}_{k,a}\bm{G}^{-1}$. 
Notice that all messages to update $q(\bm{\Phi}_{k,n})$ are now Gaussian and hence $q(\bm{\Phi}_{k,n})$ is a product of Gaussian which is also itself Gaussian with the following mean and precision matrix,
\begin{equation}\label{eq:Update_of_PHI_moments}
    \bar{\bar{\bm{\Phi}}}_{k,n}^{-1} =\hspace{-6mm} \sum_{\epsilon\in\text{Neighbourhood}}\hspace{-6mm} \bar{\bar{\bm{\epsilon}}}^{-1}, \phantom{mm} \bar{\bm{\Phi}}_{k,n} = \bar{\bar{\bm{\Phi}}}_{k,n} \hspace{-6mm}\sum_{\epsilon\in\text{Neighbourhood}}\hspace{-6mm}\bar{\bar{\bm{\epsilon}}}_{k,n}^{-1}\bar{\bm{\epsilon}}
\end{equation}
where the neighbourhood is defined as all nodes connected to  $\bm{\Phi}_{k,n}$ as shown in Fig.~\ref{fig:Baysian_graph}.
\subsubsection{Process-noise Update} The update for $q(\bm{\Lambda}_{k,a})$ can be written as \cite{bishop2007}
\begin{multline}\label{eq:surrogate_of_Lambda_a}
    \ln{q(\bm{\Lambda}_{k,a})} = \ln{p(\bm{\Lambda}_{k,a})} \\+ \sum_{n=0}^{N}\mathbb{E}_{\backslash \bm{\Lambda}_{k,a}}[\ln{p(\bm{\Phi}_{k,n}|\bm{\Phi}_{k,n-1},\bm{\Lambda}_{k,a})}]\mathds{1}_{n\neq 0} + \text{C}.
\end{multline}
By imposing a factorized gamma \gls{pdf} for $p(\bm{\Lambda}_{k,a})$, i.e., $p(\bm{\Lambda}_{k,a}) = \prod_{j=1}^4\text{Ga}(\lambda^{(j)}_{k,a}|a,\ist b)$ with shape parameter $a=\zeta/2$ and scale parameter $b=\chi/2$, then as shown in App.~\ref{app:gamma_dist_dirivation}, the proxy \gls{pdf} becomes a gamma \gls{pdf} given by
\vspace{-1mm}
\begin{align}\label{eq:post_for_Lambda_a}
    q(\lambda_{k,a,j}) = \text{Ga}\left(\lambda_{k,a,j};\hat{a},\ist\hat{b}\right)\\[-7mm]\nn
\end{align}
with $\hat{a} = (N+\zeta)/2$, $\hat{b} = (\chi + \sum_{n=0}^{N-1}[\mathbb{V}_{k,n,n+1}]_{j,j})/2$, and
\vspace{-2mm}
\begin{multline}\label{eq:def_of_V}
    \mathbb{V}_{k,n,n+1} \\= |\bm{G}^{-1}(\bar{\bm{\Phi}}_{k,n+1}-\bm{T}\bar{\bm{\Phi}}_{k,n})\rangle\langle\bm{G}^{-1}(\bar{\bm{\Phi}}_{k,n+1}-\bm{T}\bar{\bm{\Phi}}_{k,n})| \\+ \bm{G}^{-1}\bar{\bar{\bm{\Phi}}}_{k,n+1}\bm{G}^{-T} + \bm{G}^{-1}\bm{T}^{-1}\bar{\bar{\bm{\Phi}}}_{k,n}\bm{T}^{-T}\bm{G}^{-T}.
\end{multline}

\begin{algorithm}
\caption{VMP for multi object detection and tracking}\label{alg:cap}
\begin{algorithmic}
\State \textbf{Input:} Signal vector $\bm{Z}_{N}$, prior data messages $\bar{\bm{\epsilon}}_{g,0:N-1}, \bar{\bar{\bm{\epsilon}}}_{g,0:N-1}$, and cardinality $\hat{L}_N \leftarrow \hat{L}_{N-1}$ 
\State \textbf{Output:} Posterior marginals $q(\bm{\Phi}_{0:N})$, $q(\bm{\alpha}_{0:N})$, $q(\bm{\xi}_{0:N})$, $q(\bm{\Lambda}_{k,a})$ and data messages $\bar{\bm{\epsilon}}_{g,0:N}, \bar{\bar{\bm{\epsilon}}}_{g,0:N}$
\For{$k \leftarrow 1 \text{ to } \hat{L}_N$}
\State Calculate data message $\bar{\bm{\epsilon}}_{k,g,N},\bar{\bar{\bm{\epsilon}}}_{k,g,N}$ using \eqref{eq:KL_min_message}
\For{$n_I \leftarrow 0 \text{ to } N_I$}
\For{$n \in \{0\leq n\leq N : \bar{\xi}_{k,n} > 0\}$ }
\State update $q(\bm{\Phi}_{k,n})$ using \eqref{eq:Update_of_PHI_moments}
\EndFor
\State Update $q(\bm{\Lambda}_{k,a})$ using \eqref{eq:post_for_Lambda_a} and \eqref{eq:app_def_of_V}
\EndFor
\EndFor
\State Update $q(\bm{\alpha})$, and $q({\bm{\xi}}_{N})$ using \eqref{eq:update_alpha} and \eqref{eq:update_xi}.
\State For any $\bar{\xi}_{k,n}<\delta_-$ \textbf{do:} Prune track and set $\hat{L}_N \leftarrow \hat{L}_N-1$ 
\State \textbf{Using Alg.~\ref{alg:init}}: Update $q(\boldsymbol{\alpha}_N)$, $q(\bm{\xi}_N)$, $q(\boldsymbol{\Phi}_{N})$, and $\hat{L}_N$
\end{algorithmic}
\end{algorithm}

\begin{algorithm}
\caption{Initialization of new objects at time $N$}\label{alg:init}
\begin{algorithmic}
\State \textbf{Input:} Signal vector $\boldsymbol{Z}_N$, current cardinality $\hat{L}_n$, current marginals $q(\bm{\Phi}_{N})$, $q(\bm{\alpha}_{N})$, $q(\bm{\xi}_{N})$, and {threshold $\delta_+$}
\State \textbf{Output:} New marginals $q(\boldsymbol{\alpha}_N)$, $q(\bm{\xi}_N)$, and $q(\boldsymbol{\Phi}_{N})$, New cardinality $\hat{L}_\text{new}$
\State $\hat{L}_{\text{new}} \leftarrow \hat{L}_n$, $k = L_{\text{new}}$, and Continue $\leftarrow$ True.
\While{Continue}
\State $k \leftarrow k + 1$ 
\State Evaluate $f(\bm{\Phi}_{k,N}) = \langle\bm{\mu_\alpha}|\bm{\Lambda}_z|\bm{\mu}_\alpha\rangle-\ln{|\bm{\Lambda}_\alpha|}$ on a grid 
\State \hspace{3mm}of $\bm{\Phi}_N$ assuming all $\xi_{k,N}$ are 1
\State $\Check{\bm{\Phi}}_{k,N} = \text{argmax}_{\bm{\Phi}_{k,N}} f(\bm{\Phi}_{k,N}) $
\State Evaluate $\bar{\xi}_{k,N}$ using \eqref{eq:update_xi} at $\Check{\bm{\Phi}}_{k,N}$

\If{$\hat{\bar{\xi}}_{k,N}>\delta_+$}
\State $\hat{L}_{\text{new}} \leftarrow \hat{L}_{\text{new}} +1$

\State Update $q(\bm{\Phi}_{k,N})$ with $\{\bar{\bm{\Phi}}_{k,N},\bar{\bar{\bm{\Phi}}}_{k,N}\}$ using \eqref{eq:KL_min_message}
\State Update $q(\bm{\Phi}_{N}) \leftarrow q(\bm{\Phi}_{N})q(\bm{\Phi}_{k,N})$
\State Update $q(\bm{\alpha})$, and $q({\bm{\xi}}_{N})$ using \eqref{eq:update_alpha} and \eqref{eq:update_xi}.
\Else
\State Continue $\leftarrow$ False

\EndIf
\EndWhile

\end{algorithmic}
\end{algorithm}

\subsection{Algorithm Implementation}

Having derived update messages for all surrogates it is possible to update them iteratively, and as there are many loops in  the network considered (Fig.~\ref{fig:Baysian_graph}) no definitive update strategy is defined and the update strategy proposed here has not been evaluated against other procedures. Furthermore, another point of contention is the number of objects $K$ implicit in all messages considered here, as if this number is kept at the upper limit of possible objects the algorithm is computationally prohibitive. Here we propose only running all updates for the estimated number of true objects $\hat{L}_n$ and only for the time steps where the object is determined to be present. 

To achieve this, without loss of generality, we order the set of all objects such that $1\leq k \leq {L}_n$ correspond to the true objects and then further split the algorithm in two, Alg.~\ref{alg:cap}, updates the whole network for the time steps and objects with a nonzero probability of existence, to ensure convergence we iterate the messages a predetermined number $N_I$ times. Afterwards $q(\bm{\alpha}_{N})$ and $q(\bm{\xi}_{N})$ is updated and any tracks where the probability of existence falls below a threshold $\delta_-$ is pruned away by setting $\xi_{k,n}=0$. To do this in a memory efficient manner at each time step the moments of the data message is appended to $\bar{\bm{\epsilon}}_{g,0:N-1}$ and $\bar{\bar{\bm{\epsilon}}}_{g,0:N-1}$ which are data structures with variable size $\hat{L}_n\times N$. These can then be used when updating the network in subsequent time steps. Alg.~\ref{alg:init} initializes new objects to do this effectively we consider a grid for initialization, and then find the point with the highest likelihood of containing a new object by finding the point in the grid which maximizes the first two terms in \eqref{eq:update_xi} assuming $\xi_{k,n} = 1$. After which the whole of \eqref{eq:update_xi} is evaluated at this grid point, then if $\bar{\xi}_{k,n}$ is larger than some pre determined threshold $\delta_+$ the object is added. This procedure is repeated until no grid points exceed the threshold as outlined in Alg.~\ref{alg:init}.

It is worth noting the algorithm generally has low complexity with the highest complexity operation being the update \eqref{eq:update_alpha}, which is a matrix inversion in the number of tracked objects $\hat{L}_n\times\hat{L}_n$ and hence rises with $\hat{L}_n^3$, however no iterations are needed for this message to converge and hence in practice hundreds of objects may be tracked before this operation becomes prohibitive.

\section{Numerical Simulation}
\begin{table*}
    \centering
    \caption{Parameter settings}
    \begin{tabular}{ccccccccccc}
    \toprule
         $N_{T,R}$ & PRF          & $\mathbb{E}[\sigma_{RCS}]$          & G & Amplitude            & $R_{\text{max}}$    & $f_c$         & BW            & $T_{T_x}$       & $f_s$          & $\sigma_w^2$ \\ \midrule
        3         & 10 Hz & 0.05 $m^2$ & 1 & 0.53 V/m & 100 m  & 10 GHz & 20 MHz & 3.6 $\mu$s & 256 MHz & $\text{BW}\cdot k_b \cdot 290$ \\ 
        \bottomrule 
        \multicolumn{11}{l}{$k_b$ is the Boltzmann constant.}
    
    \end{tabular}    
    \label{tab:parameters}
    \vspace{-6mm}
\end{table*}

\subsection{Simulation Setup}
\begin{figure}[t]
    \centering
    \includegraphics[width=1\linewidth]{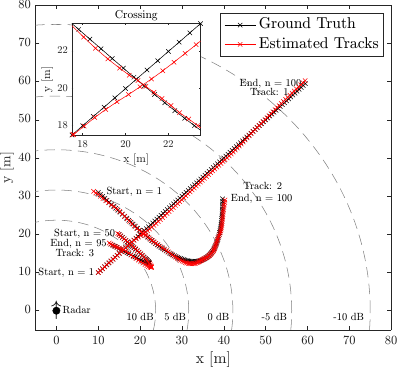}
    \caption{Example realization of scenario with three tracks. Black: ground truth; red: estimate. Dashed curves: constant mean component SNR. Track~1 starts at $(10,10)$, $ n= 1$ and moves with constant velocity $\bm{v} = 7 \begin{bmatrix}
   \cos(\pi/4) &\sin(\pi/4) 
\end{bmatrix}^T$~m/s for the entire observation time. Track~2 starts at $(10,31)$, $n=1$ and moves with constant velocity $\bm{v} = 7 \begin{bmatrix}
   \cos(\pi/4) &-\sin(\pi/4) 
\end{bmatrix}^T$~m/s for 20 time steps after which it accelerates to $\bm{v} = \begin{bmatrix}
    0 & 7
\end{bmatrix}^T$ m/s in 80 time steps. Track three starts at $(15,20)$, $n=50$ moves with constant velocity $\bm{v} = 7 \begin{bmatrix}
   \cos(\pi/4) &-\sin(\pi/4) 
\end{bmatrix}^T$ m/s for 16 time steps and then accelerates to $\bm{v}=\begin{bmatrix}
    -4.33 & 2.5
\end{bmatrix}^T$ m/s in 14 time steps and then keeps this constant velocity for 16 time steps terminating at $n=95$. Insert:  Track~1 and 2  crossing.
    }
    \label{fig:Track realization}
\end{figure}

We simulate a scenario with a radar located at the origin tracking three objects. We consider a $3\times3$ MIMO radar using
time division multiplexing . The antennas are considered isotropic, and the transmitters have half wavelength spacing while the receivers have wavelength spacing, and placed such that the resulting virtual array is a uniform linear array with half wavelength spacing.
The total observation time is 10~s with a PRF of 10~Hz yielding 100 time steps. The data is generated at each time step using \eqref{eq:signal_model_time} with $u_m(t)$ being a linear chirp and then matched filtered. The RCS follows a Swerling~3 model. Simulation parameters are listed Tab.~\ref{tab:parameters}. The tracks and their specifications of the three objects are shown in Fig.~\ref{fig:Track realization}.
It should be noted that the tracks of Objects one and two cross within 53~cm of one another at time step 22 which is considerably lower than classical range resolution of 7.5~m for this system configurations.
The survival probability is set to $p_s = 0.95$, the birth probability is set to $p_b = 10^{-8}$, the cut-off for new objects is $\delta_+ = 1-p_b$, the pruning cutoff is $\delta_- = 0.1$ the noise precision, $\bm{\Lambda}_{\bm{Z}}$, is assumed known and diagonal, and the prior precision of $\bm{\alpha}_n$ is set to $\lambda_{k,\alpha}^{(p)} = 1/\bar{\sigma}_{RCS}$, and $N_I = 100$.  To evaluate the performance the \gls{ospa} metric is used \cite{Schuhmacher2008} we use a cut-off distance of 10 m and set the order p to 2. We then average the \gls{ospa} over 900 Monte Carlo runs. When calculating $\bar{\bar{\bm{\epsilon}}}_{g,N}$ we restrict the optimization in \eqref{eq:KL_min_message} to only consider diagonal matrices with positive entries to ensure a positive semidefinite matrix.

For comparison we use a \gls{dtt} approach that pre-processes the received signal in each time step using a \gls{sbl}-based detection an localization algorithm \cite{Moederl2025}. The \gls{sbl}-based algorithm jointly detects \glspl{po} and estimates their locations (on the continuum).
The obtained detections are provided to a standard \gls{mot} method implemented in MATLAB, which uses a global nearest-neighbor approach for detection and track association. A track is confirmed once it receives at least three detections within five consecutive measurements, and it is deleted if it goes five consecutive measurements without any assigned detections, followed by constant-velocity model Kalman filter.

\begin{figure}[htbp]
    \centering
    \includegraphics[width=1\linewidth]{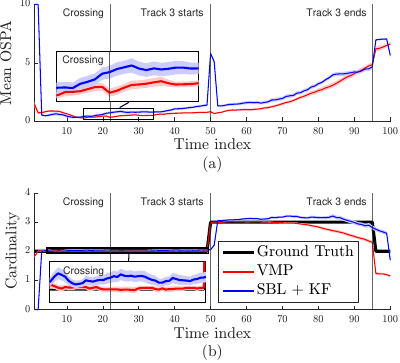}
    \caption{\textbf{(a)} shows the mean OSPA error of 900 montecarlo runs. \textbf{(b)} shows the cardinality of the ground truth, the VMP, and the SBL + KF, averaged over 900 montecarlo runs. The shaded area shows the $\pm\,3$ standard divinations of the runs. 
    }
    \label{fig:Cardinality_and_OSPA}
\end{figure}
\begin{figure}[htbp]
    \centering
    \includegraphics[width=1\linewidth]{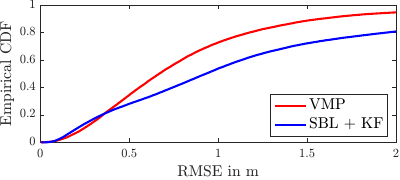}
    \caption{Empirical CDF of the RMSE error for 900 Monte Carlo runs.
    }
    \label{fig:CDF}
\end{figure}

\subsection{Results}
A realization of the VMP on the considered tracks is shown in Fig.~\ref{fig:Track realization} note that there is good tracking performance over the whole SNR region, furthermore the VMP  shows good performance both in the manoeuvrers which are different from the assumed kinematics, but also in the crossing. The mean OSPA of the tracker over 900 simulation runs can be seen in Fig.~\ref{fig:Cardinality_and_OSPA} (a), we notice large differences between the two methods around times $n=1$ and $n=50$, this is due to the confirmation rules of the SBL based algorithm as no tracks can be started before three detections are placed and hence the OSPA will converge to the cut-off distance of 10 m. This is also visible in  Fig.~\ref{fig:Cardinality_and_OSPA} (b) where the onset of true components added is delayed by $n=3$. We do not have a large cardinality or OSPA difference between the algorithms at the crossing, $n=22$, this is due to the track deletion rule "saving" the SBL, even when erroneous detections are passed to the tracker it effectively ignores them through the crossing however this also means that the disappearing track is dropped later as can be seen in Fig.~\ref{fig:Cardinality_and_OSPA} (b). Generally the VMP algorithm estimates the cardinality of the set well with very few errors before time $n=70$, the errors mainly consists of track 2 not being initialized right away as it starts in low SNR. Few spurious tracks appear for the VMP approach, whereas significantly more appear in the SBL based approach. This results in a higher average cardinality as can be seen in the insert of Fig.~\ref{fig:Cardinality_and_OSPA} (b), and  after time step 50 where the cardinality has an upwards trend. This increase in cardinality is due to track one and two being in low SNR and produce multiple spurious tracks. The cardinality error for the VMP is mainly observed as track 1 approaches the array gain limit and is hence dropped before the true track ends. It can also be seen that the VMP handles the appearing track well with no time lag in the tracks creation, likewise the track is also terminated at the right time apparent from the large drop in mean cardinality at time $n=95$.

The VMP approach handles the tracking well with a mean OSPA below one meter, and as can be seen from Fig.~\ref{fig:CDF} with 90\% of the RMSE error for tracked objects being below 1.6 meter. One feature of note is the first couple of time steps where the OSPA of the VMP raises a bit, this may be understood by the one step approach fully integrating the detection, localization, and tracking step. As the localization is informed by the tracking and the tracker is initialized by an assumption of $\bm{0}$ velocity the inertia of this prior increases the detection error before the tracker learns the true object state. 

\section{Conclusion}
The proposed direct-\gls{mot} is based on \gls{vmp} omits classical detect-then-track steps by operating on raw MIMO-radar data. In contrast to classical \gls{tbd} techniques, our approach is to model the superimposed radar signals and jointly estimating object states alongside nuisance parameters (e.g., object amplitudes, noise variance). The direct-\gls{mot} method offers advantages over traditional \gls{tbd} techniques that rely on simplified likelihood models, demonstrating robust \gls{mot} performance with reduced complexity. Our approach effectively handles low-\gls{snr} conditions and closely spaced objects.\\ \indent
Numerical results demonstrate that the proposed  approach achieves significantly lower OSPA errors than a \gls{dtt} method using super-resolution \gls{sbl} and a Kalman filter, particularly in challenging scenarios where closely spaced objects produce highly correlated measurements.

\appendices
\section{Optimization of ELBO w.r.t $q(\bm{\alpha}_n)$ and $q(\bm{\xi}_n)$}\label{app:alpha_xi_optimization}
\noindent To find the optimal choice of $q(\bm{\alpha}_n)$ and $q(\bm{\xi}_n)$ we seek to maximize \eqref{eq:ELBO_main_article}.  
We can write the ELBO as,
\begin{multline}\label{eq:app_elbo}
    \mathcal{L}(q) = \ln{G(\bar{\bm{\xi}}_n)} +\sum_{k=1}^K H(q(\xi_{k,n})) + \bar{\xi}_{k,n} g(\bar{\xi}_{k,n-1}) \\ - D_{KL}(q(\bm{\alpha})||t) + C
\end{multline}
where $g(\bar{\xi}_{k,n-1})$ is given by \eqref{eq:g_def},
\begin{equation}\label{eq:app_integral_for_G}
    G(\bar{\bm{\xi}}) = \int_{\bm{\alpha}_n} e^{\mathbb{E}_{\backslash \boldsymbol{\alpha}_n}[\ln{p(\bm{Z}_n|\bm{\Phi}_n,\bm{\xi}_n,\bm{\alpha}_n)}]+ \ln{p(\bm{\alpha}_n)}} d\bm{\alpha}_n
\end{equation}
and probability distribution,
\begin{equation}\label{eq:app_t_def}
    t(\bm{\alpha}_n;\bar{\bm{\xi}}_n) = \frac{e^{\mathbb{E}_{\backslash \boldsymbol{\alpha}_n}[\ln{p(\bm{Z}_n|\bm{\Phi}_n,\bm{\xi}_n,\bm{\alpha}_n)}]+ \ln{p(\bm{\alpha}_n)}}}{G(\bar{\bm{\xi}})}.
\end{equation}
The maximum ELBO, results from letting $q(\bm{\alpha}_n) = t(\bm{\alpha}_n;\bar{\bm{\xi}}_n)$. The expectation in \eqref{eq:app_t_def} is of  the form
\begin{multline}\label{eq:app_Loglike_iN_Terms_of_alpha}
    \mathbb{E}_{\backslash \bm{\alpha}_n}[\ln{p(\bm{Z}|\bm{\Phi}_n,\bm{\alpha}_n,\bm{\xi}_n)}] =-\langle \bm{Z}_n|\bm{\Lambda}_Z|\bm{Z}_n\rangle \\-\langle\bm{\alpha}_n|\bm{M}_n\odot\mathbb{E}_{\bm{\Phi}}[\langle\bm{S}_n^T|\bm{\Lambda}_Z|\bm{S}_n^T\rangle]|\bm{\alpha}_n\rangle \\+ \text{Re}\{\langle\bm{Z}_n|\bm{\Lambda}_Z|\mathbb{E}_{\bm{\Phi}}[\bm{S}_n^T]\bar{\bm{\xi}}\rangle|\bm{\alpha}_n\rangle\}+f(\bm{\Lambda}_Z).
\end{multline}
We then recognize $t(\bm{\alpha};\bm{\xi}_n)$ as a complex Gaussian with mean given by \eqref{eq:mu_alpha} and precision by \eqref{eq:Lambda_alpha}, which will then be the optimal distribution of $q(\bm{\alpha}_n)$. The optimal $q(\xi_{k,n})$ is found by direct optimization of \eqref{eq:app_elbo} in $\bar{\xi}_{k,n}$ as shown in \eqref{eq:update_xi}

\section{KL divergence between ${\epsilon}^{(\bm{Z}_n\rightarrow\bm{\Phi}_{k,n})}$ and ${\epsilon}_g$} \label{app:div_of_KL_divergence}
The KL divergence between
\begin{equation}
    {\epsilon}^{(\bm{Z}_n\rightarrow \bm{\Phi}_{k,n})}= e^{\mathbb{E}_{\backslash \bm{\Phi}_{k,n}}[\ln{p(\bm{Z}_n|\bm{\Phi}_n,\bm{\xi}_n,\bm{\alpha}_n)}]}
\end{equation}
and $\epsilon_g$ can be written as
\begin{equation}
    \mathcal{D}_{KL}({\epsilon}_g||{\epsilon}^{(\bm{Z}_n \rightarrow \bm{\Phi}_{k,n})}) =  -\mathbb{E}[\ln{p(\bm{Z}_n|\bm{\Phi}_n,\bm{\xi}_n,\bm{\alpha}_n)}] - H(\bm{\epsilon}_g)
\end{equation}
with the expectation w.r.t $\epsilon_g$. Carrying out the expectation yields
\begin{align}\label{eq:KL_divergence}
    &\mathcal{D}_{KL}(\bar{\bm{\epsilon}},\bar{\bar{\bm{\epsilon}}}) = C -H(\epsilon_g) \nn \\
    &\quad- 2\text{Re}\Big\{\langle\bm{Z}_n \rmv\rmv-\rmv\rmv\rmv \sum_{k'\neq k}\bar{\alpha}_{k',n}\bar{\xi}_{k',n}\bm{S}(\bar{\bm{\Phi}}_{k',n})|\bm{\Lambda}_Z|\bar{\alpha}_{k,n}\bar{\xi}_{k,n}\bm{S}(\bar{\bm{\epsilon}}_{g})\rangle\Big\} \nn \\
    &\quad + (|\bar{\alpha}_{k,n}|^2 + \bar{\bar{\alpha}}_{k,n})\bar{\xi}_{k,n}
     \bigg[ 
    \langle \bm{S}(\bar{\bm{\epsilon}}_{g}) |\bm{\Lambda}_Z|\bm{S}(\bar{\bm{\epsilon}}_{g})\rangle \nn  \\
    &\qquad + \text{Tr}\Big(\bar{\bar{\bm{\epsilon}}}_{g}\langle\nabla\bm{S}(\bm{\Phi}_{k,n})\Big|_{\bar{\bm{\epsilon}}_{g}}|\bm{\Lambda}_Z|\nabla\bm{S}(\bm{\Phi}_{k,n})\Big|_{\bar{\bm{\epsilon}}_{g}}\rangle\Big)\bigg]
    .
\end{align}
We have used the delta method to approximate the expectation of $\bm{S}(\bm{\Phi}_{k,n})$ as,
\begin{equation}
    \mathbb{E}_{\bm{\Phi}_{k,n}}[\bm{S}(\bm{\Phi}_{k,n})]\approx \bm{S}(\bar{\bm{\Phi}}_{k,n})
\end{equation}
\begin{align}
    &\mathbb{E}_{\bm{\Phi}_{k,n}}[\langle \bm{S}(\bm{\Phi}_{k,n}) |\bm{\Lambda}_Z|\bm{S}(\bm{\Phi}_{k,n})\rangle] \approx \langle \bm{S}(\bar{\bm{\Phi}}_{k,n}) |\bm{\Lambda}_Z|\bm{S}(\bar{\bm{\Phi}}_{k,n})\rangle \nn\\
    &\hspace*{4mm} + \text{Tr}\Big(\bar{\bar{\bm{\Phi}}}_{k,n}\langle\nabla\bm{S}(\bm{\Phi}_{k,n})\Big|_{\bar{\bm{\Phi}}_{k,n}}|\bm{\Lambda}_Z|\nabla\bm{S}(\bm{\Phi}_{k,n})\Big|_{\bar{\bm{\Phi}}_{k,n}}\rangle\Big)
\end{align}
with $\nabla$ denoting the gradient with respect to $\bm{\Phi}_{k,n}$. 

\section{Functional form of $q(\bm{\Lambda}_{k,a})$}\label{app:gamma_dist_dirivation}
\noindent To derive the form of $q(\bm{\Lambda}_{k,a})$ we write out the the second term of \eqref{eq:surrogate_of_Lambda_a} as,
\begin{multline}\label{eq:app_message_for_Lambda_a}
    e^{\sum_{n=1}^{N}\mathbb{E}_{\backslash\bm{\Lambda}_a}[\ln p(\bm{\bm{\Phi_{k,n}|\bm{\Phi}_{k,n-1},\bm{\Lambda}_a}})]}\\= e^{-\frac{1}{2}\sum_{n=1}^{N}Tr(\bm{\Lambda}_a\mathbb{V}^{(k)}_{n,n-1})}|\bm{\Lambda}_a|^{N/2}
\end{multline}
\begin{multline}\label{eq:app_def_of_V}    
    \mathbb{V}_{k,n,n-1} = \mathbb{E}_{\bm{\Phi}_{k,n},\bm{\Phi}_{k,n-1}}[\bm{G}^{-1}|\bm{\Phi}_{k,n}-\bm{T}\bm{\Phi}_{k,n-1}\rangle\\\times \langle\bm{\Phi}_{k,n}-\bm{T}\bm{\Phi}_{k,n-1}|\bm{G}^{-T}].
\end{multline}
 By using a factorized prior \gls{pdf}, i.e., $p(\bm{\Lambda}_{k,a}) = \prod_{j-1}^4 p(\lambda^{(j)}_{k,a})$, we may write the trace as,
 \begin{equation}
     Tr(\bm{\Lambda}_{k,a}\mathbb{V}_{n,n-1}) = \sum_{j=1}^4 \lambda_{k,a}^{(j)}[\mathbb{V}_{k,n,n-1}]_{j,j}
 \end{equation}
 resulting in \eqref{eq:app_message_for_Lambda_a} becoming completely separable in $\lambda_{k,a}^{(j)}$ as $f(\lambda_{k,a}^{(1)},\hdots,\lambda_{k,a}^{(4)}) = \prod_j f(\lambda_{k,a}^{(j)})$ with,
 \begin{equation}
     f(\lambda_{k,a}^{(j)}) = (\lambda_{k,a}^{(j)})^{N/2}e^{-\frac{1}{2}\lambda_{k,a}^{(j)}\sum_{n=1}^N[\mathbb{V}_{k,n,n-1}]_{j,j}}
 \end{equation}
which is the functional form of a gamma distribution. By carrying out the expectation in \eqref{eq:app_def_of_V} and using a gamma \gls{pdf} for $p(\lambda^{(j)}_{k,a})$, we arrive at \eqref{eq:post_for_Lambda_a}.
\bibliographystyle{IEEEtran}
\bibliography{IEEEabrv,Es_lib}

\end{document}